%% file: no-thieves-arXiv.tex
\documentclass[runningheads]{llncs}

\usepackage[main=english]{babel}
\usepackage{cite} 
\usepackage[hyphens]{url}
\makeatletter
\g@addto@macro{\UrlBreaks}{\UrlOrds}
\makeatother
\usepackage{stmaryrd}

\usepackage[T1]{fontenc}

\usepackage[
]{microtype}

\let\savevec\vec
\usepackage{amsmath}
\let\vec\savevec
\usepackage{amssymb,amsfonts}
\usepackage{xspace}
\usepackage{graphicx}
\usepackage{subcaption}
\usepackage[dvipsnames, table]{xcolor}
\usepackage{lipsum}
\usepackage{array}
\usepackage{tikz}
\usetikzlibrary{arrows.meta, positioning, fit, backgrounds, calc, shapes.geometric}
\definecolor{flabel}{RGB}{45,70,110}

\usepackage{listings}

\usepackage{booktabs}
\usepackage{multirow} 
\usepackage{wasysym}
\usepackage{todonotes}
\setuptodonotes{
  backgroundcolor=Yellow!70,
  linecolor=YellowOrange!60,
  size=\tiny
}

\usepackage{enumitem}
\setlist{nosep}

\usepackage{hyperref}
\hypersetup{
  hidelinks,
  breaklinks=true,
  bookmarksnumbered=true,
  pdfauthor={},
  pdftitle={},
  pdfkeywords={},
}
\usepackage{orcidlink}
\usepackage{xurl}
\newif\ifanonymous
\anonymousfalse

\ifanonymous
  \newcommand{\selfcite}[1]{[anonymized for review]}
\else
  \newcommand{\selfcite}[1]{~\cref{#1}}
\fi

\begin{document}


\title{It Doesn't Take a Thief: Optical-Scan Voting Systems Fail Even Without Adversaries\thanks{To appear in Proceedings of E-Vote-ID 2026, LNCS, Springer}}
\titlerunning{Optical-Scan Systems Fail Without Adversaries}

\ifanonymous
    \author{Anonymized for review}
    \institute{}
\else
    \author{Aleksander Essex\inst{1}\orcidlink{0000-0002-0228-0371}  \and
    Philip B.\ Stark\inst{2}\orcidlink{0000-0002-3771-9604}
    }
    \institute{Western University, Canada\\ \email{aessex@uwo.ca} \and
    University of California, Berkeley\\
    \email{pbstark@berkeley.edu}
    }
\fi

\maketitle

\begin{abstract}
Optical-scan voting systems and their supporting ecosystem of people, processes, and technology are fallible.
While a substantial body of work examines adversarial threats to such systems, we have encountered jurisdictions where the possibility of tabulator error is not fully internalized. Stakeholders there often find hypothetical attacks unconvincing, but some are persuaded
by real-world accounts of equipment and procedural failures.
This paper introduces a taxonomy of non-adversarial failure modes organized
into intuitive categories: recording votes on paper, reading votes from the
paper, combining votes as read into a reported outcome, and testing and verifying, all
illustrated with documented incidents.
We map common verification mechanisms against this taxonomy, identifying gaps that no paper-based audit can detect or correct, most notably failures
that compromise the trustworthiness of
the paper trail, such as giving voters the wrong ballot style (omitting contests they are eligible for, or including ones they are not),
using ballot-marking devices
to record votes, or failing to keep voted ballots
secure and organized.

\end{abstract}

\input{sections/intro-background}
\input{sections/F1-Acquisition}
\input{sections/F2-Interpretation}
\input{sections/F3-Aggregation}

\input{sections/F4-Audits}
\input{sections/disc-concl}


\ifanonymous\else
\subsubsection*{Acknowledgements.} This work was funded in part by NSERC Discovery Grant RGPIN-2026-06401 (Essex) and NSF grant SaTC–2228884 (Stark).
\fi

\bibliographystyle{splncs04}
\bibliography{refs-incidents,refs-literature}


\end{document}

%% file: sections/intro-background.tex

\section{Introduction}
\label{sec:intro}

Paper ballots are the most widely trusted medium for recording votes, for good reason: they create a durable, tamper-evident, human-readable record of voter selections independently of the method used to count the votes.
Hand counting votes from paper ballots is still common globally, providing a robust, immediate means of delivering on a universal design requirement of democratic elections to allow ``independent scrutiny of the voting and counting process $\dots$ so that electors have confidence in the security of the ballot and the counting of the votes''~\cite{unhr}.

Some jurisdictions (most notably the US) use optical-scan tabulators to speed up tabulation and
mitigate human error in vote counting.
In the United States, over 95\% of voters live in jurisdictions using paper ballots~\cite{verifiedvoting2026}.
However, a paper ballot is not a counted vote.
Between the moment a voter expresses their intent
and the moment the official results are published, votes are subject to loss, alteration, substitution, and fabrication through hardware, software, and human procedures.

While these vulnerabilities are well known among academics,
in our experience, they are not widely acknowledged by election officials
and legislators, resulting in reluctance to take countermeasures such as physical accounting for voted ballots and conducting risk-limiting audits. This paper aims to persuade such stakeholders that these measures are needed.

\subsection{Optical Scan Systems are Fallible}

There are many documented failures, with root causes ranging from unexpected `inputs' such as fold-lines in ballots being incorrectly interpreted as voter marks by inadequately maintained scanners~\cite{windham2021}, to procedural failures such as including test ballots in the election results~\cite{npr2021}, to ballot printers that are inadequate for the paper stock used~\cite{mcgregor2023}, to seemingly mundane environmental factors such as humid air at polling places causing paper jams~\cite{proPublica2018}.
Detailed incident corpora list examples dating back at least to the early 2000s~\cite{norden2010}.

A substantial body of work examines adversarial threats to voting systems, including demonstrated attacks on memory cards~\cite{feldman2007}, comprehensive security reviews~\cite{careview2007,butler2008}, and ongoing hardware vulnerability assessments~\cite{defcon27}. Although accounting for and designing to adversarial threat models is an essential activity, our experience engaging with election management bodies suggests that stakeholders are generally more receptive to real-world accounts of equipment and procedural failures than to hypothetical attack scenarios. We have even encountered institutional settings in which the very possibility of tabulator error has not been fully internalized.

To the extent that election stakeholders are the audience for this work, we believe a non-adversarial framing may be more effective at building the case for meaningful, routine post-election audits.\\

\noindent{\bf Trust and Institutional Context}. A non-adversarial framing serves both ends of the institutional trust spectrum.
In low-trust environments, it helps distinguish ordinary fallibility from unfounded fraud claims; in high-trust environments (as in Canada), it helps stakeholders internalize that trust in outcomes should rest on verifiable processes rather than assumed equipment reliability.

\subsection{Contribution and Outline}
This paper presents a systematic model of non-adversarial tabulator failure modes organized around a four-stage pipeline: vote recording, covered in \S\ref{sec:1-acquisition}; vote reading, covered in \S\ref{sec:2-interpretation}; vote aggregation, covered in \S\ref{sec:3-aggregation}; and verification and testing, covered in \S\ref{sec:4-audits}. Unlike prior taxonomies organized by root cause, we structure failures by \emph{where} they manifest in the pipeline rather than \emph{what} caused them to better align with the vantage point from which an election official actually experiences them. We map selected incidents from the United States onto this framework, demonstrating that failures have been observed at every stage of the pipeline, across vendors, jurisdictions, and election types. \S\ref{sec:implications} matches common verification mechanisms to the taxonomy, identifying gaps that no paper-based audit can cover on its own.

While prior work has systematized some sub-categories, such as optical mark detection~\cite{jones2010,bajcsy2015}, memory card reliability~\cite{davtyan2008,uconn2010}, and L\&A testing~\cite{la_testing2022}, as well as incident collections, we believe this is the first paper to propose a pipeline-structured taxonomy of non-adversarial failures spanning the full tabulation process.

\section{Background}
\label{sec:background}

The US leads global adoption of optical-scan ballot tabulators.
According to Verified Voting's 2026 polling-place equipment data~\cite{verifiedvoting2026},
96\% of US voters are in jurisdictions that use ballot tabulators: 69.5\% in jurisdictions that offer hand-marked paper ballots, with electronic ballot marking devices (BMDs) and/or direct-recording electronic touch-screens
(DREs) provided as an accessibility option, and 26.5\% in jurisdictions that require all in-person voters to use BMDs.
Only 3.9\% of US voters are in jurisdictions that use DREs as the sole in-person voting method.

Canada uses optical-scan tabulators only sub-nationally—in Ontario municipal (since the early 2000s) and provincial (since 2018) elections, and in some British Columbia, New Brunswick, and Manitoba contests—while federal elections are still counted entirely by hand. Adoption elsewhere is limited: the Philippines has used precinct-count scanners nationwide since 2010, South Korea uses central-count scanners for hand-marked ballots, and the UK has used optical scanning only occasionally (e.g., the 2004 London Assembly and 2007 Scottish Parliament elections).

There are two main optical-scan voting technologies: mark-sense and imaging scanners. Mark-sense scanners, an older technology still deployed in some US states, use an array of photoreceptors to detect reflectance at fixed positions along the ballot, interpreting marks at vote-target coordinates as votes and reading ballot-style information encoded the same way. Imaging scanners instead capture a digital image of the ballot and process it to read the votes, either from the voter's hand marks (together with encoded ballot-style information) or, for BMD printouts, from a machine-readable code; some recent systems use optical character recognition.


There are likewise two main system configurations: precinct-count (PCOS) and central-count (CCOS). In PCOS, the voter inserts the ballot into a combined scanner and ballot box. In CCOS, officials feed collected ballots (from precincts, vote centers, and mail) into a scanner at a central location. CCOS systems use high-speed scanners that are generally more accurate than the ``consumer-grade'' scanners typical of PCOS. Many US jurisdictions combine the two—for instance, PCOS for in-precinct voting and CCOS for vote-by-mail (VBM).

\subsection{Related Work}
Vulnerabilities of optical-scan ecosystems are well studied in the academic literature, which generally falls along three lines.\medskip

\begin{enumerate}
	\item {\bf Non-adversarial}: Threats to elections stemming from both potential risk and observed instances of procedural lapses and equipment failures~\cite{jones2010,careview2007,davtyan2008,smith2008}.
	\item {\bf Adversarial}: Threats to the elections stemming from cybersecurity vulnerabilities and the potential for intentionally malicious efforts to compromise integrity or ballot secrecy~\cite{defcon27,defcon25,crimmins2024dvsorder,mahmood2025busting}.
	\item {\bf Verifiability}: Efforts to make election results produced by optical-scan tabulators efficiently and independently verifiable through statistical checks like risk-limiting audits (RLAs)~\cite{rlas}, or ballot secrecy-preserving cryptographic proofs like end-to-end verifiability (E2E-V)~\cite{stii,pav,e2eviv}.
\end{enumerate}\medskip

\noindent To our knowledge, our paper is the first to propose a pipeline-structured taxonomy of non-adversarial failures covering the full tabulation stack,
but there is related work. Norden~\cite{norden2010} lists nearly 200 incidents, organized by vendor and year, rather than by failure type.
Walker et al.~\cite{la_testing2022} address logic and accuracy (L\&A) testing practices across 50 states; that work is related but focuses on a single verification mechanism rather than the full failure surface.
Jones~\cite{jones2010} categorizes optical mark-sense scanning mechanisms and image capture failure modes.
Bajcsy et al.~\cite{bajcsy2015} systematize marginal mark types as a subcategory of mark interpretation failures.
Davtyan et al.~\cite{davtyan2008} and Antonyan et al.~\cite{uconn2010} categorize memory card failures.
A NIST workshop~\cite{nist2005threats} produced a preliminary taxonomy organized around attack vectors rather than operational failure modes.
The EAC used attack trees to decompose and systematize threats~\cite{eac_threattrees2009}.

\subsection{Methodology}
\label{sec:methodology}

The incidents presented here were collected
from the academic literature on voting system security and reliability and institutional reports from the EAC, state election authorities, and civil society organizations such as the Brennan Center for Justice and Verified Voting, targeted web searches for documented incidents, discussions with domain experts, and our own direct experience auditing elections and testifying regarding contested election results.
We also used large language models as a search aid to identify candidate incidents, which we verified against primary or secondary sources.
While the incidents and analysis in this paper are drawn from the US, the failure modes we describe are inherent to the technology, not specific to any jurisdiction.

The incidents below are not an exhaustive list, but they
show that failures have occurred at every stage of the pipeline, across vendors, jurisdictions, and election types within the US. Incident-based analysis carries inherent selection bias, since failures may be undetected (especially in jurisdictions without rigorous canvasses and audits) or unreported.

\section{Tabulator Ecosystem and Failure Modes}

Optical-scan voting systems operate in a broader ecosystem of hardware, software, and human procedures that spans the full pipeline from ballot production to result certification.
Any stage can introduce errors that may be invisible to subsequent stages.
%
%
\subsection{Tabulation Pipeline and Failure Modes} 
We organize failures into four broad stages of collecting votes and reporting results:\medskip

\begin{enumerate}
  \item \textbf{F1: Vote Recording (\S\ref{sec:1-acquisition}).}
   A voter marks a ballot by hand or uses a ballot-marking device to print one; the output is a marked paper ballot, human-readable and possibly also machine-readable (e.g., a QR code). Failures present the voter the wrong contests or record something other than their expressed preferences.


  \item \textbf{F2: Vote Reading (\S\ref{sec:2-interpretation}).}
The ballot is scanned and interpreted by software configured for the election, producing a cast vote record (CVR) or simply incrementing contest vote counters. Failures record or use the selections differently from what a reasonable human observer would read from the ballot.


  \item \textbf{F3: Vote Aggregation (\S\ref{sec:3-aggregation}).}
The votes as-read, possibly partially aggregated, are transmitted to a central election management system (EMS) and combined into official results. Failures turn otherwise correct interpretations into incorrect totals and outcomes.


  \item \textbf{F4: Verification and Testing (\S\ref{sec:4-audits}).}
Checks before the election (logic and accuracy testing) and after it (ballot accounting, pollbook reconciliation, chain-of-custody and seal checks, post-election audits) can prevent or recover from F1--F3 failures, unless the marked ballots themselves are untrustworthy, or the tests are missing, poorly designed, or poorly executed.

\end{enumerate}\medskip

\noindent Vulnerabilities of optical-scan systems extend well beyond the tabulator hardware and software: the election definition software (F2) and the EMS (F3) are external systems whose failures are indistinguishable from tabulator failures in the final result.\\

\noindent{\bf Relationship to Security Definitions.} \emph{Software independence} (SI) requires that an undetected error in a voting system's software cannot cause an undetectable change in an election outcome~\cite{rivest2008}, and is a formal requirement of the VVSG 2.0~\cite{eacvvsg2}.
Optical-scan systems can be software-independent: hand-marked paper ballots provide a human-readable record independent of the software used to count them.
An election is \emph{defensible}~\cite{appel2020ballot} if the election official can provide affirmative public evidence that the reported outcomes
are correct, despite any errors or malfeasance that might have occurred.
\emph{Evidence-based elections}~\cite{starkWagner12,appelStark20} provide affirmative public evidence that elections were decided correctly
by combining demonstrably trustworthy vote records with risk-limiting audits (F4).
SI, defensibility, and being evidence-based are properties of a run of the complete system
(an election), not just the hardware.
The primary failure modes (F1--F3)
indicate what an audit should protect against.

\subsection{Terminology}

We distinguish among three related notions that are often conflated:\medskip

\begin{enumerate}
	\item {\bf Voter intent}. The candidate or option the voter intends to select.
    Voter intent is known only to the voter.
	\item {\bf Expressed preference}.
    The selection the voter {\em actually makes} by marking a ballot by hand or indicating a selection on a BMD using the touchscreen or an audio interface.
    Expressed preferences differ from voter intent if the voter errs in expressing their intent~\cite{appel2020ballot}.
	\item {\bf Marked selection}.
    The human-readable selection physically recorded on the paper ballot.
    In some BMD systems, the tabulator reads votes
    encoded in QR codes or barcodes, not
    the human-readable votes, but we consider the human-readable text to be the marked selection.
\end{enumerate}\medskip

\noindent
In the context of this discussion, the {\em marked selection} is the ground truth against which tabulator output is evaluated.
In rare cases, the marked selection is ambiguous to a reasonable human observer (for instance, the marked selection was ambiguous in only a handful of ballots in the 2008 Franken–Coleman state-wide recount~\cite{mnrecount2008}).
The distinction between (2) and (3) arises in the case of ballot-marking devices.
With hand-marked ballots (HMPBs), the voter's expressed preference is the marked selection.
With BMD-printed ballots, the expressed preference may differ from the marked selection as the result of an F1 vote recording failure, as the DeKalb County 2022 incident in the next section illustrates.


%% file: sections/F1-Acquisition.tex

\section{Vote Recording (F1)}
\label{sec:1-acquisition}


\subsection{Components}
The correct output of vote recording is a paper ballot that accurately reflects the voter's expressed selections in every contest the voter is eligible to vote in (and no other contests), in human-readable form and optionally also machine-readable.
F1 begins with a printed blank ballot or a BMD configured for the election and ends with a marked paper ballot:\medskip

\begin{description}[nosep,leftmargin=1em,labelindent=0em,labelsep=0.35em]
\item[Ballot printing and/or BMD configuration.]
  BMDs sometimes fail to function entirely, resulting in `denial-of-voting.' Jurisdictions such as the state of Georgia and Los Angeles County, CA, that require all in-person voters to use BMDs are especially vulnerable to such failures.
  Both HMPB and BMDs are subject to \emph{ballot presentation} failures: the printed ballot or the BMD interface may omit or include some contests or candidates erroneously.
\item[BMDs.] BMDs translate the voter's expressed preferences in the presented contests into selections on a printed ballot.
  The voter then has the opportunity to inspect the printed ballot before inserting it into the scanner.
  All-in-one BMDs (e.g., ES\&S ExpressVote~XL) may (at the voter's instigation)
  deposit the printout directly into a ballot box, bypassing the opportunity to review the printed
  selections.
\item[HMPB.] With hand-marked paper ballots, the paper record begins as an accurate record of
   the voter's expressed preferences in the contests presented on the ballot.
\item[Curation of cast ballots.] The cast ballots are the physical ground truth for post-election audits and recounts.
  Maintaining secure custody of the ballots
  is an artifact-integrity matter: losing, adding, damaging, or altering cast ballots undermines the trustworthiness of the election, regardless of whether other processes functioned correctly.

\end{description}

\begin{table}[t]
\centering
\small
\setlength{\tabcolsep}{5pt}
\renewcommand{\arraystretch}{1.2}
    \begin{tabular}{l
                    >{\raggedright\arraybackslash}p{2.8cm}
                    >{\raggedright\arraybackslash}p{2.0cm}
                    c
                    >{\raggedright\arraybackslash}p{3.8cm}}
    \toprule
    {\bf Ref.} &{\textbf{Incident}} & \textbf{Component} & \textbf{Type} & \textbf{Outcome} \\
    \midrule
    \cite{faussetEtal20} &
    Georgia, 2020 &
    BMD & P, A &
    BMDs did not function \\
    \cite{fowler20} &
    Georgia, 2020 &
    BMD & P, A &
    BMD does not display all candidates in a contest \\
    \cite{horsely26}
    & Louisville, KY, 2021--2026 &
    HMPB & P, A &
    Voters assigned to the wrong precinct, presented the wrong contests \\
    \cite{dekalb2022a,dekalb2022b}
    & DeKalb County, GA, 2022
    & BMD & P & Incorrect ballot def. Selections assigned to the wrong candidate \\
    \cite{spotlightPA2023}
    & Northampton County, PA, 2023
    & BMD & P & Expressed preferences swapped for two candidates \\
    \bottomrule
    \end{tabular}
\caption{Vote recording incidents (F1). P = Process failure;\enspace A = Artifact integrity failure.}
\label{tab:f1}
\end{table}

\subsection{Selected Incidents}

Table~\ref{tab:f1} summarizes selected F1 incidents. We briefly elaborate on three cases.\medskip

\noindent{\bf Voters given incorrect ballot style in
Louisville, KY, 2021--2026.}
\cite{horsely26}
A clerical error in assigning voters to precincts resulted in giving voters ballots that omitted some contests they were eligible to vote in.
Some errors that result in giving voters the wrong collection of contests are recoverable and some are not.
In general, if contests (or candidates) are omitted, the error is unrecoverable because voters' selections were never solicited and recorded.
Depending on details, it can be possible to recover from errors that erroneously add a contest to ballots,
for instance, if the error occurs for every ballot of a given style.
\medskip

\noindent{\bf BMD fails to display some candidates in Georgia, 2020}
A bug in Dominion voting system software (not a configuration error) caused BMDs to fail to display about half the candidates in a US Senate special election~\cite{fowler20}.
The accidental discovery of the bug during L\&A testing
triggered Dominion to issue a software patch
that was deemed a \emph{de minimis} change by the
US Election Assistance Commission\footnote{%
See also \url{https://www.eac.gov/voting-equipment/engineering-change-orders/dvs-100714}, last accessed 26 July 2026.
}
and therefore exempted from additional testing---even though the original bug had slipped through VVSG certification testing.\medskip

\noindent{\bf BMDs in DeKalb County, GA, 2022.}
Following the withdrawal of a candidate in the Democratic primary for DeKalb County Commission District~2, a ballot definition error in Dominion BMD touchscreens caused votes intended for one remaining candidate to be recorded as votes for another~\cite{dekalb2022a,dekalb2022b}. In this instance, the error was introduced when the BMD committed the voter's marked selections to paper (the tabulators themselves interpreted and counted ballots correctly). This case illustrates a characteristic of BMD failures: because the paper ballot reflects what the BMD printed rather than what the voter selected, software independence holds only if the voter verified the printed ballot before casting it, which studies have shown most voters do not perform in practice~\cite{bernhard2020,appel2020ballot}.

%% file: sections/F2-Interpretation.tex
\section{Vote Reading (F2)}
\label{sec:2-interpretation}


\subsection{Components}
F2 failures include mark-detection errors, stray marks or paper folds read as votes, ballot-definition errors, discrepancies between a vote's human-readable and barcode/QR encoding, and incorrect application of vote-validity rules. The F2 pipeline scans a ballot and uses ballot-definition information to interpret the scan, creating a (possibly ephemeral) cast vote record that is stored or used to increment contest vote counters. This interpretation depends on a chain of logical mappings that must be mutually consistent across the ballot definition, the tabulator, and the EMS:\medskip

\begin{description}[nosep,leftmargin=1em,labelindent=0em,labelsep=0.35em]

\item[Scanning.] The scanner ingests the ballot under the prevailing environmental conditions.
  Process failures include mechanical feed problems such as jams, double-feeds, and skipped ballots or batches of ballots~\cite{proPublica2018,jones2010,stark2024}, and optical failures such as sensor degradation, dirty scanner heads~\cite{windham2021}, image skew or distortion, print-through from the opposite side of a ballot sheet,
  insufficient resolution and insufficient image bit depth.
  Humidity can disrupt feed mechanisms and degrade image quality~\cite{jones2010}.
  Sensor contamination accumulates over time if maintenance is inadequate~\cite{windham2021}.

\item[Ballot definition (process).] The ballot definition process encompasses authoring and configuration to present the correct collection of contests with the correct choice options to each voter and to read marked ballots.
For HMPB, this generally includes mapping physical target positions in the ballot geometry to candidate identifiers in the tabulation system, depending on ballot-style information encoded on the printed ballot.
Process failures include mapping selections to the wrong contest or candidate, or applying the wrong voting rules to the contest.
Because a single mapping error affects every ballot of a given style, these failures tend to be systematic and are often undetectable from aggregate results.

\item[Ballot definition file.] The ballot definition file is a digital artifact that carries the ballot geometry and candidate/position mappings into each tabulator. For interpretation to be correct, three things must be mutually consistent: (1) the physical positions of vote targets on the printed ballot, (2) the definition file loaded on the tabulator, and (3) the version used by the EMS to decode CVRs. Any pairwise mismatch produces a failure. A last-minute ballot revision may be propagated to tabulators and the EMS, but leave already-printed ballots inconsistent with the new definition. Alternatively, a revision may update the EMS but not all tabulators, as in the 2020 Antrim County incident~\cite{halderman2022}.

\item[Mark interpretation.] The mark interpretation process evaluates the digitized ballot image against the ballot definition to determine which ovals contain valid votes. Process failures arise from threshold calibration errors. Over-sensitive configurations count stray marks as votes, and under-sensitive ones miss valid marks. Process failures also arise from physical ballot artifacts such as fold-lines or smudges that mimic filled ovals, and from sensor-to-sensor variation that causes identical marks to be evaluated differently across machines~\cite{windham2021}.

\item[Cast vote record (CVR).] The CVR is a data artifact produced by mark interpretation, encoding the tabulator's per-ballot vote assignments for downstream aggregation for some optical-scan systems. Artifact integrity failures include file corruption between the tabulator and the EMS and unauthorized modification of CVRs prior to aggregation.
\end{description}

\begin{table}[t]
\centering
\small
\setlength{\tabcolsep}{5pt}
\renewcommand{\arraystretch}{1.2}
\begin{tabular}{l
                >{\raggedright\arraybackslash}p{2.8cm}
                >{\raggedright\arraybackslash}p{2.0cm}
                c
                >{\raggedright\arraybackslash}p{3.8cm}}
\toprule
\textbf{Ref.} & \textbf{Incident} & \textbf{Component} & \textbf{Type} & \textbf{Outcome} \\
\midrule
    \cite{mcgregor2023}
    & Maricopa County, AZ, 2022
    & Scanning & P & Timing marks too faint to read \\[2pt]
    \cite{proPublica2018}
    & New York City, NY, 2018
    & Scanning & P & Paper jams (humidity) \\[2pt]
    \cite{ncsbe2018}
    & Wake County, NC, 2018
    & Scanning & P & Ballots rejected (humidity)\\[2pt]
    \cite{md2016}
    & Maryland, 2016
    & Scanning & P & Double-feeds. Scratched lens caused false overvotes \\[2pt]
    \cite{cambria2024}
    & Cambria County, PA, 2024
    & Scanning & P & Printing error made ballots unreadable \\[2pt]
    Fig.~\ref{fig:sf2024}
    & San Francisco County, CA, 2024
    & Scanning & P & Scanner erases light marks \\[2pt]
\cite{halderman2022}
  & Antrim County, MI, 2020
  & Ballot definition file & A & Mismatched definition. Votes mapped to incorrect candidates \\[2pt]
\cite{windham2021}
  & Windham, NH, 2020
  & Mark interpretation & P & Fold-lines incorrectly interpreted as votes \\[2pt]
\cite{appel2023}
  & Los Angeles County, CA, 2023
  & Mark interpretation & P & Stray marks outside ovals read as overvotes\\
\bottomrule
\end{tabular}
\caption{Vote reading incidents. P = Process failure;\enspace A = Artifact integrity failure.}
\label{tab:f2}
\end{table}

\subsection{Selected Incidents}

Table~\ref{tab:f2} summarizes selected F2 incidents. We elaborate on four, with two of them illustrating failures at opposite ends of the F2 pipeline: sensor-level mark interpretation (Windham) and ballot definition files (Antrim).\medskip

\noindent{\bf Scanning in Maricopa County, AZ, 2022.}
Ballot-on-demand printers failed to maintain sufficient fuser temperature, causing timing marks to print too faintly for Dominion ICP tabulators to read~\cite{mcgregor2023}. Ballots were later re-tabulated on central-count scanners, an outcome made possible only because paper ballots were retained and a recovery procedure was in place. The incident illustrates both the sensitivity of scanning to upstream printing conditions and the importance of contingency procedures for F2 faults.\medskip

\begin{figure}[t]
  \centering
  \begin{subfigure}[b]{0.48\textwidth}
    \includegraphics[width=\linewidth]{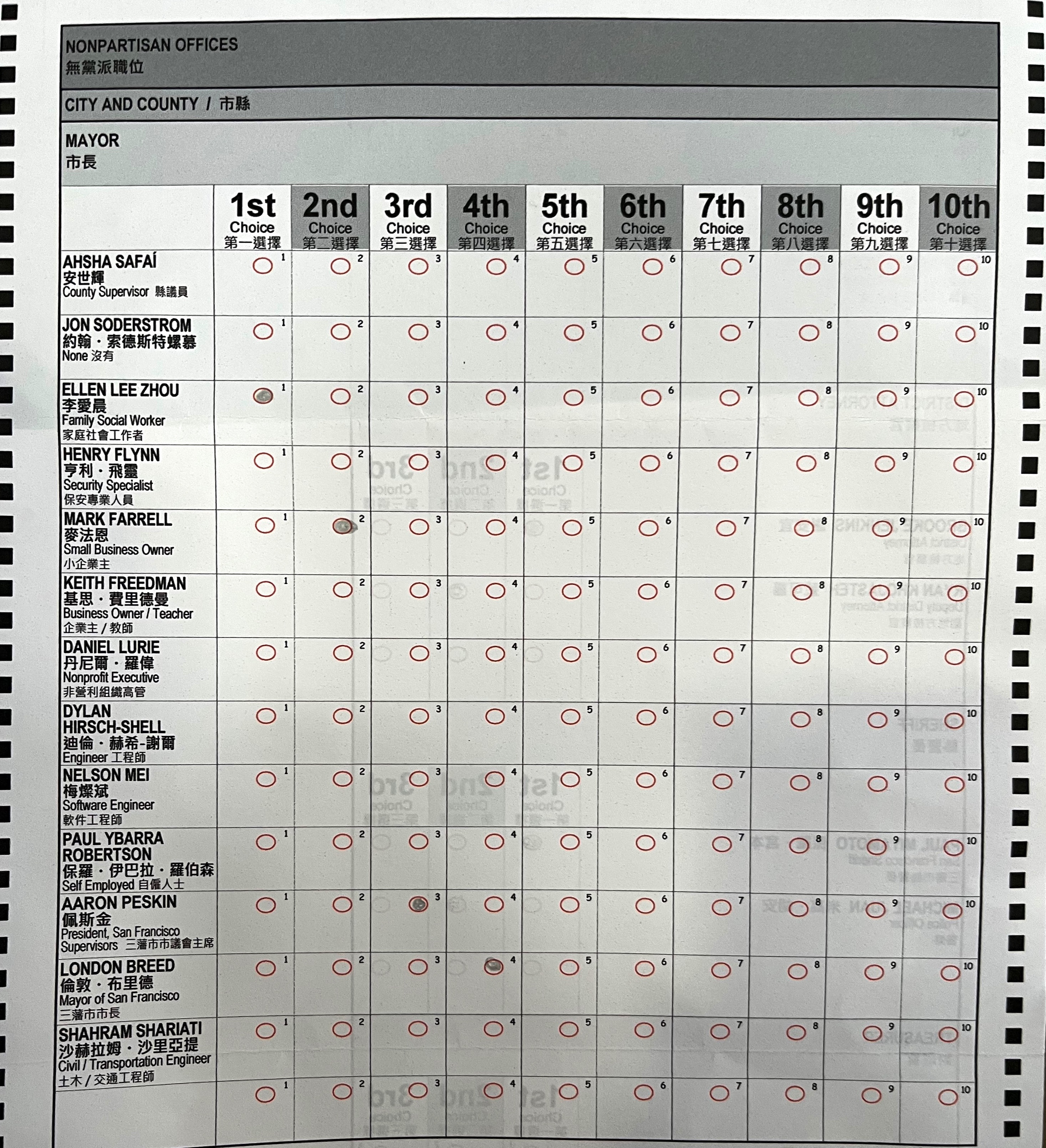}
    \caption{Author-taken photograph of a ballot.}
  \end{subfigure}
  \hfill
  \begin{subfigure}[b]{0.48\textwidth}
    \includegraphics[width=\linewidth]{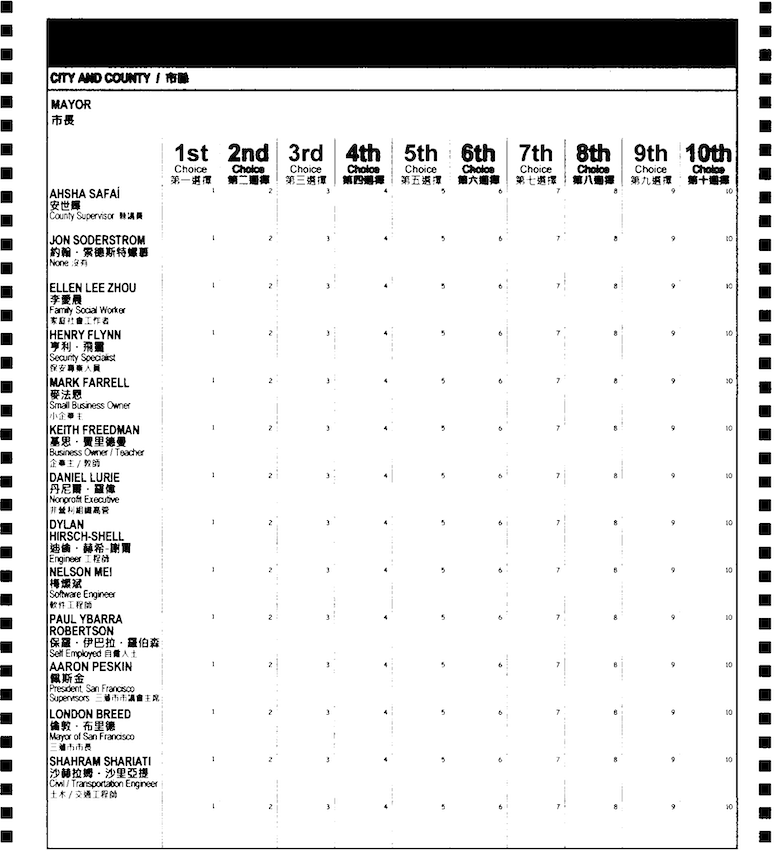}
    \caption{What the Dominion scanner captured.}
  \end{subfigure}

  \caption{\textbf{Example F2 Failure.} Two conflicting images of a ballot cast in San Francisco in November 2024. {\bf (a)} Photo taken by one of us (Stark) during the post-election audit showing valid voter-made marks. {\bf (b)} Image created by the scanner failing to capture marks. Observe the print-through from the other side is visible in (a). Lowering the image's black threshold could result in print on the back side being interpreted by the scanner as a marked oval on the front side.}
 \label{fig:sf2024}
\end{figure}

\noindent{\bf Scanning in San Francisco, CA, 2024.}
  Figure~\ref{fig:sf2024} shows a digital picture
  of a voted ballot and a Dominion optical-scan system scan of the same ballot, from an
  election in San Francisco in November 2024.
  The voter's marks are clear in the
  picture, but there is no trace of the marks in the scanned image because the Dominion scanner records the image in black-and-white (not grayscale or color) and the mark is below the threshold
  that renders as black.\medskip


\noindent{\bf Mark Interpretation in Windham, NH, 2020.}
AccuVote mark-sense scanners misinterpreted fold-lines on absentee ballots as voters' marks~\cite{windham2021}.
A letter-folding machine had been repurposed to handle a surge in absentee ballots, and the resulting creases intersected candidate vote ovals, creating erroneous votes and overvotes.
Inconsistent maintenance left the scanners with varying sensor contamination, so the same creased ballots were read differently across machines.\medskip

\noindent{\bf Ballot Definition File in Antrim County, MI, 2020.}
A last-minute ballot revision added a write-in blank to one township's ballot, causing the Dominion EMS to reassign sequential vote-target IDs across all ballot styles. Many tabulators were not reprogrammed with the updated definition, leaving them operating with a stale ballot definition file. When the EMS decoded the resulting memory cards against the new IDs, votes were mapped to the wrong candidates~\cite{halderman2022}. A full hand recount confirmed the correct result, illustrating the three-way consistency requirement. Notably, all three independent records disagreed: the poll tapes reflected the tabulator's totals under the old definition, the EMS totals reflected decoding under the new definition, and the paper ballots retained the ground truth. The divergence illustrates why retaining paper ballots {\em and} printed tally tapes is important. The hand count corrected the outcome, while the tape-vs-EMS discrepancy identified where in the pipeline the failure occurred.

%% file: sections/F3-Aggregation.tex
\section{Vote Aggregation (F3)}
\label{sec:3-aggregation}
%
%
%
\subsection{Components}
F3 failures tend to be systematic: a single procedural lapse or software error can affect entire batches, precincts, or jurisdictions. The F3 pipeline carries per-ballot CVRs through accumulation and transmission to produce jurisdiction-wide totals:\medskip

\begin{description}[nosep,leftmargin=1em,labelindent=0em,labelsep=0.35em]
  \item[Vote accumulation.] The tabulator
  accumulates votes across all ballots processed in a session.
  Process failures include arithmetic errors, counter overflow, and software bugs that cause votes to be dropped or double-counted.
  One infamous category of aggregation failure involves {\em negative} vote totals, which can arise from arithmetic faults or corrupted counters~\cite{norden2010}.

  \item[Results artifact (memory card).] The vote totals results artifact is typically implemented as a removable memory card, which carries per-tabulator totals and CVRs from the polling place to the EMS. Memory cards can be lost, uploaded multiple times, initialized with corrupt data, or simply never uploaded. The 2020 Georgia hand recount identified both failure modes: approximately 5{,}900 ballots across four counties were found uncounted because memory cards had never been transferred to the EMS, while in Fulton County some ballot cards were included in the tally multiple times in the original count and in a machine recount due to procedural failures~\cite{stark2024}.

  \item[Election management system (EMS).] The EMS aggregates results from all tabulators into a jurisdiction-wide result. Process failures include software bugs that drop or duplicate batches, misconfigured aggregation logic, and (as in the  NYC 2021 incident~\cite{npr2021}) failure to clear test data before tabulation. Because the EMS process operates on results from multiple tabulators simultaneously, failures here tend to have the widest scope of any component in the pipeline.
\end{description}

\begin{table}[t]
\centering
\small
\setlength{\tabcolsep}{5pt}
\renewcommand{\arraystretch}{1.2}
\begin{tabular}{l
                >{\raggedright\arraybackslash}p{2.8cm}
                >{\raggedright\arraybackslash}p{2.0cm}
                c
                >{\raggedright\arraybackslash}p{3.8cm}}
\toprule
\textbf{Ref.} & \textbf{Incident} & \textbf{Component} & \textbf{Type} & \textbf{Outcome} \\
\midrule
\cite{norden2010}
  & Volusia County, FL, 2000 & Results artifact & A & Memory card reported negative vote totals \\
\cite{uconn2010}
  & Connecticut, 2007--09
  & Results artifact & A & Up to 15\% of memory cards contained random data after battery replacement \\[2pt]
\cite{stlucie2012}
  & St.\ Lucie County, FL, 2012
  & Results artifact & A & Procedural errors caused double-scanned and unscanned ballots \\[2pt]
\cite{npr2021}
  & NYC Mayoral Primary, 2021
  & EMS & P & Test ballots included in tabulation, skewing results \\[2pt]
\cite{sullivanNJ2023,shorenewsNJ2023}
  & Monmouth County, NJ, 2022
  & EMS & P & Omitted software patch led to double-counted votes \\[2pt]
\cite{stark2024}
& Fulton County, GA, 2020 & Results artifact & A & Ballots counted multiple times or omitted due to inadequate tracking \\[2pt]
\cite{alameda-abc7-2022}
& Alameda County, CA, 2022 & EMS & P &
Ranked-choice voting rules for wrong
jurisdiction used. Wrong winner reported \\
\bottomrule
\end{tabular}
\caption{Vote Aggregation incidents (F3). P = Process failure;\enspace A = Artifact integrity failure.}
\label{tab:f3}
\end{table}

\subsection{Selected Incidents}

Table~\ref{tab:f3} summarizes selected F3 incidents. We elaborate on two cases that illustrate both process and artifact integrity failures at this stage.\medskip

\noindent{\bf Results Artifact in Volusia County, FL, 2000.}
In this landmark incident, a memory card in Precinct 216 reported {\em negative} $16{,}022$ votes for Al Gore~\cite{norden2010}, which the EMS did not range-check. The erroneous total contributed to premature reporting, calling Florida for George W. Bush.  The VVSG 2.0 now explicitly requires that electronic devices detect and prevent the accumulation of negative votes~\cite{eac_vvsg2}, a requirement that did not exist at the time.\medskip

\noindent{\bf EMS Process Failure in Monmouth County, NJ, 2022.}
In July 2022, an ES\&S technician reinstalled election management software on Monmouth County's server to resolve a network connectivity issue, but omitted a software patch designed to prevent duplicate ballots~\cite{shorenewsNJ2023}.
Some votes in four municipalities were double-counted, leading to the wrong candidate being certified until a recount identified the error some months later~\cite{sullivanNJ2023}.
The incident illustrates how the absence of reconciliation checks between precinct-level totals and EMS aggregates allowed the error despite system certification.

%% file: sections/F4-Audits.tex
\section{Verification and Testing (F4)}
\label{sec:4-audits}


\subsection{Components}
Unlike F1--F3, where the failure surface is defined by hardware, software, and data artifacts, the F4 failure surface involves the \emph{design} of the verification process, including what it checks and how thoroughly it covers these failure surfaces relative to the paper ballot ground truth.
It also involves the \emph{execution} of the audit by human auditors.
For instance, post-election audits involve retrieving ballot sheets,
manually interpreting the votes on those sheets, and entering the interpretation into
the audit software. The F4 pipeline checks results against independent artifacts:\medskip

\begin{description}[nosep,leftmargin=1em,labelindent=0em,labelsep=0.35em]
  \item[Logic and Accuracy (L\&A) testing.] L\&A testing
    is the primary pre-election check in most
    jurisdictions. A test deck of ballots is processed by each
    tabulator, and the tabulator results are compared against the ground-truth totals of the deck.
    Process failures in L\&A testing arise from insufficient
    test deck design, such as not testing all relevant
    vote combinations, ballot styles, or environmental factors.
    A fifty-state review found that L\&A implementations vary dramatically in scope and rigorousness across jurisdictions~\cite{la_testing2022}.

    \item[Inventory and chain-of-custody procedures.] Maintains a complete accounting of physical ballots, memory cards, and tally tapes between pipeline stages, and protects artifact integrity in transit. Process failures include not physically accounting for ballot cards (relying on the voting system to know how many cards were cast),
    inadequate tracking of which memory cards have been used and whether they have been uploaded, which ballot boxes have been sealed and by whom, and which tabulators have been accounted for. Detection of missing artifacts requires reconciliation against an independent expected count (e.g., poll books and other participation records, ballot accounting, and VBM processing records). The Georgia 2020 hand recount uncovered approximately 5{,}900 uncounted ballots in four counties attributable to memory cards never uploaded to the EMS~\cite{stark2024}, which proper reconciliation ought to have caught.

  \item[Post-election audits.] Post-election audits are
    the primary mechanism by which F1--F3 failures can be detected
    and corrected after an election. Their effectiveness depends on the integrity of the paper trail and the design of the audit.
    If the record of the votes is incomplete, incorrect, or adulterated (some F1 failures), little can be done; an audit could even confirm a wrong result.
    An audit that checks only a fixed
    percentage of precincts or that relies on the same software or scans used for tabulation provides little assurance.
    Risk-limiting
    audits (RLAs)~\cite{stark2010} of a trustworthy paper trail provide statistical guarantees of correcting incorrectly reported outcomes.
    Relatively few US jurisdictions conduct RLAs, and those that do generally audit at most a few contests, which may create a false sense of confidence in the overall election because auditing some contests cannot confirm any contest that was not audited.

    \item[Software and firmware integrity checks.]
    Some jurisdictions require verification that tabulators are running certified software before an election, typically through cryptographic hash comparison. Like L\&A testing, these are pre-election checks, but they establish which software was installed rather than whether any reported outcome is correct. Process failures include hash scripts that report success regardless of whether a valid hash file is present~\cite{ess_hash2020} and self-attestation in which a device reports its own hash.

%
%

\end{description}

\begin{table}[t]
\centering
\small
\setlength{\tabcolsep}{5pt}
\renewcommand{\arraystretch}{1.2}
\begin{tabular}{l
                >{\raggedright\arraybackslash}p{2.8cm}
                >{\raggedright\arraybackslash}p{2.0cm}
                c
                >{\raggedright\arraybackslash}p{3.8cm}}
\toprule
\textbf{Ref.} & \textbf{Incident} & \textbf{Component} & \textbf{Type} & \textbf{Outcome} \\
\midrule
\cite{windham2021,la_testing2022}
  & Windham, NH, 2020
  & L\&A testing & P & Fold-line failure not caught under operational conditions \\[2pt]
\cite{ess_hash2020}
  & ES\&S hash verification, 2020
  & Software integrity check & P & Hash script reported success with missing hash file \\[2pt]
\cite{spotlightPA2023}
  & Northampton County, PA, 2023
  & Post-election audit & P & Barcode correct, but hand count of human-readable text showed incorrect outcome \\[2pt]
  \cite{stark2024}
  & Fulton County, GA, 2020
  & Post-election audit/handcount & P & Audit tally sheets omitted from audit results\\
\bottomrule
\end{tabular}
\caption{Verification and Testing incidents (F4). P = Process failure.}
\label{tab:f4}
\end{table}

\subsection{Selected Incidents}
\label{sec:f4-incidents}
Table~\ref{tab:f4} summarizes selected F4 incidents. We elaborate on two.\medskip

\noindent{\bf L\&A Testing in Windham, NH, 2020.} Pre-election L\&A testing did not catch the Windham fold-line failure (\S\ref{sec:2-interpretation}) because the test deck did not include folded ballots~\cite{windham2021}. This is a characteristic L\&A failure: successful verification under sanitized test conditions, failure under operational ones.


\noindent{\bf Audit Design Limitations in Northampton County, PA, 2023.} In this case, the supposed paper ballot ground truth was itself internally inconsistent: the barcode used for tabulation was correct while the human-readable summary was not, a divergence undetectable without independently decoding the barcode.


%% file: sections/disc-concl.tex
\section{Implications for Audits}
\label{sec:implications}

The F1--F4 taxonomy provides a structured basis for evaluating which verification mechanisms can detect which failure classes. A recurring theme of the incident analysis is that no single verification mechanism covers the full pipeline. Meaningful post-election audits, therefore, must take the form of a layered system of checks whose combined coverage spans the first three failure categories.




%
%

\begin{table}[t]
\centering
\caption{Coverage of failure modes by verification mechanism.
Symbols address detection, not recoverability. \CIRCLE~= detects. \LEFTcircle~= partially detects (see notes).
\Circle~= does not detect. L\&A = {\em logic and accuracy testing}.
BA/CoC = {\em ballot accounting and chain-of-custody reconciliation}. BP-RLA = {\em ballot-polling RLA}.
BC-RLA = {\em ballot-comparison RLA}. Batch = {\em batch-comparison audit}.
HR = {\em hand recount}.}\medskip

\label{tab:coverage}
\begin{tabular}{@{}llccccccc@{}}
\toprule
& Failure mode & L\&A & BA/CoC & BP-RLA & BC-RLA & Batch & HR \\
\midrule
\multirow{3}{*}{\textbf{F1}\phantom{.}}
& Ballot printing/BMD config. & \LEFTcircle\rlap{$^{a}$} & \LEFTcircle\rlap{$^{b}$} & \Circle & \Circle & \Circle & \Circle \\
& BMD marking fidelity        & \Circle & \Circle & \Circle & \Circle & \Circle & \LEFTcircle\rlap{$^{c}$} \\
& Curation of cast ballots    & \Circle & \CIRCLE & \Circle & \Circle & \Circle & \Circle \\
\midrule
\multirow{5}{*}{\textbf{F2}\phantom{.}}
& Scanning                    & \LEFTcircle\rlap{$^{a}$} & \Circle & \LEFTcircle\rlap{$^{d}$} & \LEFTcircle\rlap{$^{d}$} & \LEFTcircle\rlap{$^{d}$} & \CIRCLE \\
& Ballot definition (process) & \LEFTcircle\rlap{$^{a}$} & \Circle & \CIRCLE & \CIRCLE & \CIRCLE & \CIRCLE \\
& Ballot definition file    & \LEFTcircle\rlap{$^{e}$} & \Circle & \CIRCLE & \CIRCLE & \CIRCLE & \CIRCLE \\
& Mark interpretation         & \LEFTcircle\rlap{$^{a}$} & \Circle & \CIRCLE & \CIRCLE & \LEFTcircle\rlap{$^{f}$} & \CIRCLE \\
& CVR integrity               & \Circle & \Circle & \Circle & \CIRCLE & \LEFTcircle\rlap{$^{f}$} & \CIRCLE \\
\midrule
\multirow{3}{*}{\textbf{F3}\phantom{.}}
& Vote accumulation           & \LEFTcircle\rlap{$^{g}$} & \Circle & \CIRCLE & \Circle & \CIRCLE & \CIRCLE \\
& Results artifact (memory)   & \Circle & \CIRCLE & \LEFTcircle\rlap{$^{d}$} & \LEFTcircle\rlap{$^{d}$} & \LEFTcircle\rlap{$^{d}$} & \CIRCLE \\
& EMS aggregation             & \Circle & \Circle & \CIRCLE & \Circle & \CIRCLE & \CIRCLE \\
\bottomrule

\end{tabular}

\vspace{2pt}
{\footnotesize
$^{a}$ Only what the test deck checks (e.g.\ not folded ballots, marginal marks,
untested styles). $^{b}$ Reconciliation against participation records; omitted
contests are unrecoverable. $^{c}$ Only if printed text and its encoding
disagree. $^{d}$ Only failures large enough to alter outcome. $^{e}$ Only if
re-run on every tabulator after the last definition change. $^{f}$ Whole batches
only. $^{g}$ In-tabulator only; EMS out of scope.
}
\end{table}

\subsection{What Each Verification Mechanism Covers}

Table~\ref{tab:coverage} maps six verification mechanisms against the F1--F3
sub-components of Sections~\ref{sec:1-acquisition}--\ref{sec:3-aggregation}. L\&A coverage is bounded by test deck design and fails under
operational conditions the deck omits (\S\ref{sec:4-audits}, motivating Recommendation~\ref{rec:latest}).
Ballot accounting and chain-of-custody reconciliation is the only mechanism that
detects missing, duplicated, or never-uploaded artifacts; the audits assume its
result rather than establishing it, which is why the F1 curation and F3
results-artifact rows depend on it (Recommendation~\ref{rec:chain}). Ballot-polling RLAs
detect only outcome-altering failures~\cite{stark2010}. Ballot-comparison RLAs check CVR
integrity between tabulator and EMS, but are blind to errors in combining the
votes as-read, where the CVRs faithfully match the ballots. Batch comparison
detects only whole-batch failures. Hand recounts give the strongest F2--F3
coverage but are typically triggered only by margin thresholds or discovered
discrepancies. This creates a paradox in jurisdictions where tabulators are used without RLAs, since the
evidence needed to trigger a recount may be demonstrable only by a recount.
Every Canadian jurisdiction using tabulators is caught in it, and none has
signalled a move toward RLAs.

The F1 rows that matter most for the paper trail are uncovered. Votes
never solicited (wrong ballot style) or never recorded (BMD unavailable) cannot
be recovered from the ballots cast, and no paper-based mechanism detects a BMD
printout that misstates the voter's expressed preference, since all of them
treat the printed text as ground truth. Northampton County 2023 (\S\ref{sec:f4-incidents}) is the exception, and only because the text and the barcode disagreed (Recommendations~\ref{rec:bmdlimit} and~\ref{rec:bmdver}).

\subsection{Recommendations}

The following recommendations follow directly from our taxonomy, incident corpus, and verification coverage analysis in the preceding sections:\medskip

\begin{enumerate}
  \item \label{rec:universal}\textbf{Post-election audits should be universal and mandatory.} Our analysis demonstrates that failures have occurred across all pipeline stages, across all major vendors, and in jurisdictions with well-resourced election administrations. Importantly, the absence of a {\em detected} failure in a given election is not evidence that the outcome was correct.

  \item \label{rec:taxonomy}\textbf{Audit design should be informed by the full failure taxonomy.} Jurisdictions should evaluate their audit procedures against each of F1--F3 explicitly and identify gaps in coverage.

  \item \label{rec:latest}\textbf{L\&A test decks should be designed to exercise known failure modes.} Test decks should include folded ballots, ballots with marginal marks, and all vote combinations for each contest, including split-ticket and RCV patterns. As commonly practiced, L\&A checks none of these, nor mark-density thresholds or sensor sensitivity.

  \item \label{rec:chain}\textbf{Inventory reconciliation should be completed before certification.} The Georgia 2020 memory card failures and the Monmouth County double-counting were detectable by reconciling precinct-level records and ballot counts against independent expected counts (poll books, VBM processing). This reconciliation should be a required pre-certification step in every jurisdiction, rather than left to ad hoc post-election discoveries.

  \item \label{rec:bmdlimit}\textbf{BMD deployments require explicit acknowledgment of fragility and audit limitations.}
  Jurisdictions using BMDs for all voters (especially all-in-one designs where the voter cannot inspect the printout) should explicitly recognize that (1)~Election day failures may prevent voting altogether, so HMPB backup ballots are required, and (2)~Audits cannot detect errors in recording votes originating in BMDs, so genuine RLAs are impossible.
  These facts should be publicly acknowledged.

  \item \label{rec:bmdver}\textbf{New verification mechanisms for BMD F1 failures are needed.}
  Ensuring BMD printouts reflect voters' expressed preferences is an open, evidently intractable problem \cite{starkXie22}.
\end{enumerate}

\section{Conclusion}
\label{sec:conclusion}


Optical-scan elections have failed repeatedly without any adversary, in ways that are well documented. No stage, vendor, or jurisdiction is immune. The F1–F4 taxonomy organizes those failures by where in the pipeline they manifest. Hand-marked paper ballots kept secure and accounted for throughout make it possible to catch outcome-changing errors and correct wrong outcomes—that is, to have 
software independence. But the possibility of checking is not the same as {\em actually} checking.  Every contest should be audited. Audits should be designed with the full failure taxonomy in mind. In low-trust settings, audits of a trustworthy paper trail answer unfounded claims of manipulation; in high-trust settings, they replace assumed reliability with demonstrated correctness.